# Controlled redistribution of vibrational population by few-cycle strong-field laser pulses

**William A. Bryan,*[a,d] C. R. Calvert,[b] R. B. King,[b] J. B. Greenwood,[b] W. R. Newell[c] and I. D. Williams[b]**

a) Department of Physics, Swansea University, Swansea, SA2 8PP, UK. E-mail: w.a.bryan@swansea.ac.uk;
b) Centre for Plasma Physics, School of Mathematics and Physics, Queen's University Belfast, Belfast, BT7 1NN, UK
c) Department of Physics and Astronomy, University College London, WC1E 6BT, UK
d) STFC Rutherford Appleton Laboratory, Harwell Science and Innovation Campus, Didcot, Oxon, OX11 0QX, UK

The use of strong-field (i.e. intensities in excess of $10^{13}$ Wcm$^{-2}$) few-cycle ultrafast (durations of 10 femtoseconds or less) laser pulses to create, manipulate and image vibrational wavepackets is investigated. Quasi-classical modelling of the initial superposition through tunnel ionization, wavepacket modification by nonadiabatically altering the nuclear environment via the transition dipole and the Stark effect, and measuring the control outcome by fragmenting the molecule is detailed. The influence of the laser intensity on strong-field ultrafast wavepacket control is discussed in detail: by modifying the distribution of laser intensities imaged, we show that focal conditions can be created that give preference to this three-pulse technique above processes induced by the pulses alone. An experimental demonstration is presented, and the nuclear dynamics inferred by the quasi-classical model discussed. Finally, we present the results of a systematic investigation of a dual-control pulse scheme, indicating that single vibrational states should be observable with high fidelity, and the populated state defined by varying the arrival time of the two control pulses. The relevance of such strong-field coherent control methods to the manipulation of electron localization and attosecond science is discussed.

## 1 Introduction

Quantum mechanics describes internal molecular dynamics in terms of the amplitude and phase of vibrational wavepackets, which are coherent superpositions of states and contain all information about the associated populations and phases. In the energy or frequency domain, the quantum beating is observable by resonant photonic processes; in the temporal domain, this interference results in characteristic time-varying motion. Ultrafast laser systems generating near-infrared (NIR) pulses with durations of hundreds of femtoseconds allowed the first observation of such wavepackets[1,2] with significant applications in chemical dynamics, opening up the field of femtochemistry. Vibrational wavepackets have been observed in a range of systems, often initiated by optical pumping with an ultrafast laser pump pulse and observed by fragmenting the molecule with a similar probe or dump pulse. Recent advances have allowed such wavepacket motion to be resolved in individual molecules[3] through the application of single-molecule detection schemes.

A range of coherent control strategies have been demonstrated whereby temporal or spectral shaping is applied to laser pulse to modify the launch or evolution of vibrational wavepacket motion, thus altering internal states of the molecular system under study.[4] Generally, such modifications require optical coupling between states, and complex spectral or temporal shaping is required to populate pre-defined final state.[5]

State-of-the-art ultrafast laser technology allows access to the few-cycle regime, where typically the near-infrared (NIR) photon energies correspond to a electric field period of around 2.7 fs. Ultrabroadband Ti:sapphire oscillators with dispersion control optics and chirped pulse amplification in temperature-controlled crystals produce (NIR) pulses with a duration of the order 20–30 fs over a bandwidth of >30 nm.[6] By spectrally broadening through self-phase modulation in a gas-filled hollow fibre,[7–9] the bandwidth is extended to hundreds of nanometers which allows subsequent compression durations as short as 3.3 fs[10] with few-mJ pulse energies and kHz repetition rates. Such pulses are basis for attosecond XUV pulse generation via high-order harmonic generation.[11]

The vibrational periods of the lightest and simplest molecules $H_2^+$ and $D_2^+$ are 13 and 20 fs respectively. NIR few-cycle pulses are therefore perfectly suited for imaging vibrational wavepackets as a pulse duration shorter than the vibrational period is readily achievable. Using an interferometrically stable pump–probe configuration and reflection focusing tight enough to generate an intensity of the order $10^{14}$ Wcm$^{-2}$ allows ultrafast strong-field imaging of vibrational motion. The wavepacket is generated by tunnel ionization of the neutral molecules,[12,13] projecting the ground state wavefunction in the neutral molecule onto all vibrational states in the molecular ion. The ionization rate varies as a function of internuclear separation,[14] hence in long (i.e. > 50 fs) laser pulses the subsequent ionization and fragmentation dynamics is dominated by enhanced ionization at large internuclear separation,[12] facilitated by dissociative wavepacket dispersion. In a few-cycle pulse, the wavepacket initiated by tunnel ionization is well localized at a small internuclear bond, a result of the favorable temporal conditions.

Following the launch of the vibrational wavepacket by a few-cycle pump, a similar strong-field probe pulse is applied to the $H_2^+$ or $D_2^+$ molecular ions, initiating fragmentation via photodissociation or Coulomb explosion.[15,16] Measuring the distribution of kinetic energy release allows the wavepacket shape to be imaged via the well-known potential energy surfaces. As is frequently the case in molecules, the vibrational wavepacket in $H_2^+$ or $D_2^+$ was predicted to dephase due to anharmonicity of the potential surface, and rephases or revives when all components interfere in phase.[17] Such vibrational wavepackets were observed in hydrogenic molecular ions by a number of groups: Cocke and co-workers observed the initial dephasing of hydrogenic wavepacket,[18] Moshammer, Ullrich and co-workers then studied $D_2^+$ observed dephasing and revival.[19] The authors and co-workers imaged similar behaviour in $HD^+$, observing dephasing and revival[20,21] and explored the coherence of a $D_2^+$ wavepacket by stretching the pump and probe pulses.[22]

At the high intensities employed in ultrafast studies of vibrational wavepackets using strong-field pulses, a significant rotational impulse is applied by the field induced torque on the molecule.[23] The time the torque is applied for is far shorter than the characteristic period of rotational motion, hence such intense pulses launch a rotational wavepacket through impulsive alignment, generating a coherent superposition that continues to evolve once the initial pulse is turned off. Rotational wavepackets have been observed in hydrogenic molecules[24,25] by observing fragmentation which is modulated by the revival of the wavepacket, and are well described by recent theory.[26]

It might intuitively appear that the conditions associated with generating vibrational and rotational wavepackets are very similar if not identical, however recent investigations by the authors indicate that either a rotational wavepacket in $D_2$ or a vibrational wavepacket in $D_2^+$ is generated.[22,25] This is not to say that the processes are mutually exclusive, however in recent experiments there is no evidence for coherent rovibrational excitation. We suspect this is the result of interactions occurring in different volumes: at focus, the ionization rate is high, hence a vibrational wavepacket is generated before rotation gets underway. Off focus, the ionization rate is much lower hence impulsive alignment can cause rotation. Alternatively, the frequency resolution is limited so as not to be able to resolve rovibrational components which in the future could be rectified by extending the pump–probe delay to the picosecond timescale. As detailed in ref. 25, rotational and vibrational contributions can be isolated by bandpass filtering. Also, see ref. 48 for a recent review of studies of vibrational and rotational wavepackets in hydrogenic molecular ions.

Time- and frequency-domain measurements of wavepackets in molecules facilitated by techniques such as two-dimensional Fourier-transform spectroscopy[27] allow the inter- and intranuclear transfer of energy to be resolved on femtosecond and nanometer scales, which have proven to be short enough to resolve complex biochemical reactions. A very promising alternative has been proposed for small molecules exposed to strong-field few-cycle pulses: quantum beat imaging employs a pump–probe configuration identical to earlier systems, with the pump causing ionization to launch a (ro)vibrational wavepacket and the probe fragments via Coulomb explosion.[12] By measuring the pump–probe delay dependent kinetic energy release from $D_2^+$ and applying a filtered Fourier transform, a section of the laser-dressed potential energy surface has been recovered from the experimental wavepacket.[28] This proof-of-principle measurement led to the theoretical extension of quantum beat imaging to a range of pulse durations and intensities[29] and multiple spatial dimensions.[30–32] Recent experimental observations of vibrational wavepackets in electronically complex diatomics[33,34] will lead to this method being extended to larger systems. The real potential of this technique lies in the ability to resolve the laser-modified potential energy surfaces without prior knowledge.

As with coherent control investigations, once it has been established that a vibrational (or indeed rovibrational) wavepacket has been generated and imaged, it is natural to try to modify its evolution. Spectral shaping unavoidably leads to an increase in pulse duration, which defeats the purpose of employing few-cycle pulses. Recent theoretical predictions indicate a strong-field few-cycle pulse applied at the correct time can heavily perturb a bound electron orbital leading to the transfer of vibrational population as the nuclei adjust nonadiabatically to the rapidly varying electronic environment.[35–37] This process can be treated as a dynamic Raman or Stark process.[38] In the former, multiphoton coupling between ground and excited states as wavepacket oscillates causes a bond length (hence time) dependent redistribution; the latter is a polarization of the molecular orbital by the dipole force leading to a time-varying distortion of potential surfaces. The nuclear wavepacket then propagates on the modified potential, and the diabaticity of the process causes population transfer. In both cases, changing intensity, wavelength and intensity of control pulse influences population transfer. Impressive experimental evidence for time-dependent manipulation of a vibrational wavepacket has been presented by Niikura and co-workers,[39,40] however only a limited portion of the timeseries is reported.

In the rest of this paper, we briefly review a recent theoretical model which predicts the strong-field modification of a vibrational wavepacket. This efficient quasi-classical model allows the influence of the distribution of intensity within a focal volume to be explored. We then discuss in detail the experimental facilitation of control of a vibrational wavepacket in $D_2^+$. Finally, we extend the quasi-classical model to a dual-control pulse scheme, again investigating how changing relative delays and intensity allows range of vibrational states to be populated with varying fidelity. Finally, the prospect for manipulating electron localization as a nuclear wavepacket evolves under external control is discussed.

## 2 Theoretical model

The scheme for vibrational wavepacket control by a strong-field few-cycle pulse is sketched in Fig. 1, and is comparable to those discussed theoretically. The pump pulse initiates tunnel ionization of $D_2$, populating a coherent superposition of vibrational states in the electronic ground state ($1s\sigma_g$) of the $D_2^+$ molecular ion. The manipulation pulse which induces a dipole force is applied after some delay during which the vibrational wavepacket has evolved in time, as shown in Fig. 1b for the solution of the time-dependent Schrodinger equation and the equivalent quasi-classical trajectory model (QCM), Fig. 1c. Finally, the subsequent evolution of the modified wavepacket is mapped by a probe pulse, via photodissociation (PD) of the molecular ion.

In the QCM (see Bryan *et al.*[41] for details), the $1s\sigma$ $v = 0$ ground state of the neutral $D_2$ precursor is projected onto the available electronic states in $D_2^+$ ($1s\sigma_g$ ground, $2p\sigma_u$ dissociative) and $D^+ + D^+$ (Coulomb explosion) states by tunnel ionization[42] and the relative populations found by numerically solving the resulting coupled differential rate equations. Stable $D_2^+$ ions are generated over a narrow pump intensity range of $4 \times 10^{13} < I_{pump} < 1.1 \times 10^{14}$ Wcm$^{-2}$, and a range of vibrational states ($0 < v < 24$) populated. The wavepacket is approximated as a classical ensemble of particles moving on the 1ssg potential, and allowed to evolve in a Newtonian manner.

As the orbital configuration defines the potential surface constraining the nuclear motion, by polarizing the electrons with a strong field, the nuclear dynamics can be controlled. In the present work, the key mechanism is nonadiabatically altering the nuclear environment by coupling the electric field associated with the control pulse to the nuclear motion via the transition dipole and the Stark effect. The former is a transient dipolar polarization created by the interaction of the electromagnetic field of the control pulse with the molecular orbital (most influential at small internuclear separations), the latter is the direct shifting and distortion of the molecular potential by the applied field (dominant at large internuclear separations). It is assumed that the mass of the nuclei makes the polarization of these positive particles negligible.

We model the time-varying modification of the classical ensemble trajectories during the control pulse, which adjust in amplitude, frequency (hence v-state) and phase offset. The control pulse is applied around the time of the first return of the wavepacket to its inner turning point, between the dashed lines on Fig. 1(b and c) as the wavepacket is still well localized at this time. Comparing the perturbed and unperturbed ensemble trajectories allows the final v-state populations and relative phases to be deduced. Rather than having to quantify the coupling between the $1s\sigma_g$ and $2p\sigma_u$ states during photodissociation in the probe pulse, we use the critical R-cutoff method, which has been proven to be a highly efficient approximation.[43] It is this predicted PD yield as a function of pump–probe and pump–control delay that will be compared directly to experimental yields. Following photodissociation which releases a kinetic energy dependent on the vibrational states populated, the fragment ions ($D^+$ and D) travel along the polarization direction of the probe, with the charged fragment recorded in an ion spectrometer.

A vital consideration for strong-field control is that any attempt to manipulate the wavepacket should not destroy either the coherence of the system or photodissociate the molecule. The photon energy and bandwidth of the control pulse could introduce complex inter-state couplings which must be quantified. For a spectrally well-defined control the manipulation operation should also be carefully tailored: readily ionized molecules are not able to withstand the same level of polarization by the field as a molecule with a high ionization potential, hence the intensity and duration of the control pulse is a key experimental parameter.

Fig. 2 illustrates the importance of having a well-quantified distribution of laser intensity and imaging a subsection of the focal volume. The Huygens–Fresnel diffraction integral is numerically solved for an experimental system, which has typical parameters of $\lambda$ = 790 nm, hollow-fibre internal diameter 250 $\mu$m, recollimated 1 metre from exit, focused into the spectrometer with f = 50 mm spherical mirror, $1/e^2$ beam radius D = 1.5 mm, total propagation distance from source to focal volume L = 5 m, and is comparable to a number of experimental systems world-wide. The laser beam profile on exit of our hollow fibre (see later) is reasonably described by the predicted $EH_{11}$ hybrid mode. Propagation through our optical system unavoidably introduces diffraction which makes the focal volume non-Gaussian, the relevance of which to atomic ionization studies has been well documented,[44,45] hence is expected to be highly influential when experimentally realizing strong-field control. Taking $\log_{10} I(x,y,z)$ of the resulting three-dimensional intensity distribution, then calculating the histogram of relative volume contributions allows a straightforward application of the QCM results. This 'focal histogram' (Fig. 2a) also allows the relative radius of the PD-yield detector to be investigated.

The beam waist at focus is approximately 10 mm, generating a pump pulse peak intensity approaching $10^{15}$ Wcm$^{-2}$. If the ion detector radius is 1 mm, the full variation of intensity over the focal volume is seen, with significantly more volume at low intensities: over the intensity range of interest the focal histogram varies by two orders. As the radius of the detector is reduced, the subsection of the focal volume available exhibits a massive change in distribution of relative intensity. As the detector radius approaches 125 μm, the focal histogram is predicted to be constant over the intensity range of interest.

The result of operating the QCM at a peak pump intensity between $10^{13}$ Wcm$^{-2}$ and $1.58 \times 10^{14}$ Wcm$^{-2}$, a pump–control pulse intensity ratio of 3:1, and durations of 6 fs FWHM is presented in Fig. 2(b to h). The QCM returns the initial and final vibrational populations, predicted PD yield as a function of pump–probe delay and work done as a function of vibrational population. Note that the results are not normalized for relative volume. At low intensity, Fig. 2(b to e), there is minimal evidence for the influence of the control pulse, as the predicted PD yield is essentially that of the unperturbed wavepacket, i.e. a pump–probe configuration without a control pulse. As the intensity of the pump is further increased for the same pump–control ratio, Fig. 2(f to h), the initial population distribution is observed to shift to a higher v-state and the distribution of final vibrational states increasingly modified. There is also a clear shift in the predicted PD yield structure, and the work done or by the control pulse increases. If an experimental measurement was attempted with a 2 mm detector, the essentiality unperturbed wavepacket motion shown in Fig. 2(b to e) would swamp the control-modified motion in Fig. 2(f to h), and would most likely be unobserved.

By reducing the diameter of the detector, the focal volume imaging is altered, as evidenced from the focal histogram modification. As the detector diameter is reduced, the subset of the focal volume imaged shifts in favour of the higher intensities. As the pump–control intensities are set by the relative energies of the two pulses, a more prominent variation of the predicted photodissociation yield is seen as more work is done on the wavepacket. At a detector radius of 125 μm, the intensity histogram is flat, and as we've taken $\log_{10}I(x,y,z)$, this results in a heavy bias in favor of highest intensity. As demonstrated in the following section, this makes strong-field control of the vibrational wavepacket experimentally observable. Furthermore, as the photodissociation of $H_2^+$ and $D_2^+$ is strongly peaked around the probe polarization direction,[12] we further increase our ability to observe the influence of the control pulse.

## 3 Experimental demonstration

The recently published first demonstration of strong-field control of a vibrational wavepacket (see Bryan *et al.*[46]) requires three precisely-timed intense laser pulses to interact with an hydrogenic molecule in the gas phase. Linearly polarized pulses from a Ti:sapphire FemtoLasers CompactPro HP (30 fs, 800 nm, 1 mJ at 1 kHz) were focused into a ~1 m long, 250 mm internal diameter hollow fibre with a pressure gradient of argon (~$10^{-5}$ to 2 bar), increasing the bandwidth from 30 nm FWHM to ~140 nm. Eight chirped mirror reflections minimized the group-delay dispersion (GDD) producing a 10.2 fs pulse as measured in an all-reflective FROG with an error of G = 0.009.

The pump, control and probe pulses are interferometrically derived from the output of the hollow fibre as illustrated in Fig. 3; all beam splitters (Femtolasers low dispersion parts OA135, OA037 and OA200) introduce minimal GDD. The input energy into the nested Mach–Zehnder interferometer was 240 μJ; following optical losses, the P:C:P pulse energies were 32:11:35 μJ, polarized perpendicular: parallel:parallel with respect to the ion spectrometer axis. The relative delay between the three pulses is independently controllable with two high-precision translation stages (Newport MFA-CC). To establish P:C:P pulse synchronization, an autocorrelation signal was monitored as the P-C and P-P delays are independently scanned and temporal satellites observed from P-C and P-P cross correlation. We estimate the temporal overlap (hence P-C-P delay) uncertainty is ~300 attoseconds, derived from the translation state resolution and reproducibility and pixel size in the autocorrelator CCD camera.

An effusive jet of room temperature $D_2$ is introduced into the interaction region of an ion time-of-flight mass spectrometer (TOFMS)[20,22,25,46–48] containing an aperture of radius 125 μm (cf Fig. 2), which essentially reduces this to a 1D problem given the prevalence for photodissociation along the laser polarization direction, hence the spectrometer is only sensitive to molecules aligned to within a few degrees of the probe polarization direction. Following precompensated transmission through a fused silica window, the P-C-P pulse sequence was reflection focused into the interaction region of the TOFMS by a spherical f = 50 mm silver mirror. A room temperature beam of $D_2$ was crossed effusively with the laser focus, and the generated $D_2^+$ and $D^+$ ions electrostatically separated; the kinetic energy of fragmentation was also measured for the $D^+$ ions. Fig. 4 shows the focal histogram for the experimental set-up, confirmed by measuring the spatial distribution of atomic ionization in argon;[44] differences between Fig. 2a and 4 are due to the finite diameter of the optics and vacuum entrance window being considered sequentially. This distribution of intensities is employed when integrating the QCM over the focal volume, thus allowing a very accurate comparison with the measured photodissociation yields. Ions arriving at the end of the TOFMS drift tube struck a pair of microchannel plates, and the resulting electron cascade collected on a solid anode. The induced voltage was monitored by a digital phosphor oscilloscope (DPO, Tektronix DPO-7254B), sampled at 2.5 GS/s; the DPO also drives both delay lines.

Following data collection, the $D^+$ fragmentation yield was integrated over 0–1 eV, the full range of photodissociation energies.[12] A Butterworth bandpass filter was applied to the PD yield as a function of pump–probe delay to suppress highfrequency noise and the low-frequency contribution from the $D_2$ rotational wavepackets.[22,25] The parameters of this filter (high and low pass frequencies and sampling frequency) were optimized by superimposing two representative sine waves of varying frequencies and adding random noise of variable amplitude, then filtering to recover the higher frequency wave. While there is a frequency overlap at the extremes of the vibrational (16 to 47 THz, peaking at 42 THz) and rotational (5 to 20 THz, peaking at 12 THz) spectra, the asymmetric distribution of population across the vibrational and rotational states[25] allowed the Butterworth filter to accurately recover the high frequency component even with 100% noise modulation.

The post-filtering PD yields associated with the evolution of the vibrational wavepacket as a function of probe delay for different control delays are presented in Fig. 5, where the control and probe delays are always referenced to the pump pulse at t = 0. As the control delay is varied between 18 fs and 32 fs, the experimental PD yield exhibits subtle but statistically significant variations: the oscillatory structure is observed to increase in amplitude, cycle-averaged yield level and periodicity. Qualitatively, this demonstrates that the wavepacket amplitude in the higher vibrational states has increased, indicative that the molecular orbital distortion by the control pulse is redistributing vibrational population. A variation of oscillation shape is also observed, from saw-tooth (18 fs) to more sinusoidal (22 fs) to almost plateau (28 fs), finally to large amplitude features well-separated from t = 0 (32 fs). Considering that the average period of $D_2^+$ is around 24 fs, such rapid variations in structure reveals that shifting the P-C delay by 2 fs steps dramatically changes the bound wavepacket.

After isolating the vibrational signatures in the experimental data, the volume integrated QCM results are overlapped. The similarities in shape, amplitude, cycle-average yield offset and periodicity of the theoretical prediction and measured PD yield indicates the coherence of the vibrational wavepacket is retained throughout the control pulse. This is the first experimental demonstration of not only the subsequent evolution of a modified $D_2^+$ vibrational wavepacket in an ultrashort strong-field pulse, but also the first evidence that the vibrational population and phases of the superposition can be modified in the ultrafast strong-field regime in a predictable manner.

The vibrational populations corresponding to the best-fit signal-volume-integrated PD yields (Fig. 5) are presented in Fig. 6. It should be noted that these results are not derived directly from the data presented in Fig. 5, rather are simulations using the known experimental parameters, hence the phrase ‘predicted’ is employed. Sources of uncertainty are defining the zero delay time (estimated as 300 as from a linear delay calibration) and the range of active intensities (estimated as better than $8 \times 10^{12}$ $\mathrm{Wcm^{-2}}$ for all three pulses). These uncertainties are combined in the QCM to produce upper and lower bounds indicated by the thickness of the theoretical prediction. Clearly, uncertainty in intensity and either delay will result in a significant variation of the QCM outcome, indicated in Fig. 6.

The initial vibrational population caused by the pump pulse is indicated (green line). The final vibrational population distribution (black line) is observed to vary significantly as the control delay is scanned. When the control pulse is applied at 18 fs, the final state distribution is driven coherently towards the lowest *v*-states: *v* = 0–3 contain almost all the population. As the control delay is increased, the lowest *v*-states are depopulated relative to the 18 fs distribution, particularly the *v* = 0, 1 states. At a control delay of around 24 to 26 fs a significant population exists in *v* = 3 and 4, and this population shoulder is observed to move to a higher *v*-state as the control delay increases up to ~30 fs, at which point the distribution of *v*-states becomes structured. At a control delay of greater than 26 fs, the lowest *v*-states are depopulated with respect to the initial conditions. The shifting of the vibrational population from the low to high *v*-states as the control delay is increased is compatible with our earlier qualitative discussions of the periodicity and oscillatory structure.

By integrating the product of the force applied and distance travelled by each ensemble element during the control pulse, the work done on the wavepacket can be predicted as a function of vibrational state. As shown in Fig. 6, at a P:C delay of 18 fs, the majority of vibrational states work on the field-modified potential, which is responsible for the skewing of the population distribution to lower *v*. As the P:C delay is increased, the distribution of work done shifts to higher *v*, as the control pulse acts more efficiently on more highly lying *v*-states. For P:C delays above 26 fs, the work done is seen to oscillate rapidly which shifting to higher vibrational state, causing the evolution in the structure of the vibrational population.

The phases of the QCM trajectories are distorted by the control pulse; while this classical phase is not strictly a quantum phase, it is an intuitive indicator of the ensemble motion. For all final *v*-states in $D_2^+$, the distorted trajectories are compared with the unperturbed trajectories (Fig. 1c), and a population-weighted mean calculated, as presented in Fig. 6. The maximum phase distortion as a function of final vibrational state follows the maximum population change, with phase changes of the order $\pi$ radians exhibited. Importantly, a large phase change is possible irrespective of the state population, opening the possibility for storing quantum information in the vibrational phase.

Fig. 7 shows volume-integrated vibrational population and phase matrices illustrating the redistributive action of the control pulse. During the operation of the QCM, all perturbed trajectories are compared to the unperturbed trajectories, resulting in an $n \times n$ matrix between 0 and 1 indicating the best fit to the final *v*-state and an $n \times n$ phase matrix between 0 and $2\pi$. In both cases, $n$ is the number of vibrational states considered, here $n$ = 24. The transfer function is converted to a vibrational population matrix by scaling by each row by the initial vibrational population. Were no control pulse present, the population matrices would be a diagonal line; population above the diagonal is up-shifted in vibrational state. Depending on the control delay and *v*-state, up- and down-shifting is clear: summing the matrix horizontally results in the initial distribution of states; summing vertically generates Fig. 6. The asymptotic tendency at high *v* is a result of the strongest influence of the control pulse at the largest internuclear separations. The empty area at a final state of $v > 13$ indicates that all trajectories in these states photodissociate. The opposite effect is responsible for the lowest-lying *v*-states being minimally influenced. A population above $10^{-5}$ is the threshold for generating the phase matrices. Note that the phase is represented in terms of fractions of an oscillation in the final *v*-state.

## 4 Modelling multiple control pulses

To demonstrate the feasibility of more extensive control of the vibrational state of a molecule with strong-field pulses, we now extend the QCM to two control pulses with independent delays. The proposed technique is distinct from traditional coherent control methods as, rather than applying one shaped pulse, we propose that multiple few-cycle impulsive actions on the wavepacket at well-timed intervals can achieve a useful fidelity of final state. This approach has been discussed for solution to the TDSE for unique double control pulse scenario,[36] however here we further the state of the art by systematically investigating the control landscape. Such a study is a demonstration of the efficiency of the QCM, as solving the TDSE repeatedly would take a prohibitive long time.

The QCM is readily modified to include an additional control pulse as the ensemble trajectory motion is modelled at each time-step, allowing a control field of arbitrary complexity. As with the single-control case, the QCM is run out to 750 fs in 50 as steps, and returns the photodissociation yield, phase and population matrices and work done as a function of vibrational state. Operating the QCM 20,000 times requires 42 h on a standard-performance PC (Intel 2.4 GHz Core 2 Duo P8600, 4 GB RAM, Microsoft Windows 7). Scaling to more complex molecules will require an increase in computing power: each populated electronic state will have to be included, and for tri- and polyatomics, the QCM will have to be extended to cover each active degree of freedom. If resonant control pulses are applied, interstate coupling will have to be introduced and a more complete representation of rovibrational dynamics is necessary. Finally, as discussed earlier, volume integration is clearly necessary. While such modelling is significantly more advanced than the current discussion, the problem is perfectly suited to parallelization methods.

Experimentally, deriving additional control pulses from one ultrafast pulse is nontrivial as repeated interferometric splitting of the laser output leads to significant loses, leading to a drop in the peak intensities available. An alternative to additional nested interferometers is the use of a focusing optic divided into independently movable annuli, however for large delays the spatial overlap of the resulting focal volumes would degrade. Furthermore, such optics introduce significant diffraction,[44,45] however the resulting distortion to the pulse wavefronts may compensate for focal walk-off. Another method for introducing independent delays is the azimuthal rotation of glass plates; again, by dividing into annuli, different spatial elements of the pulse can be delayed but this introduces a delay-dependent group delay dispersion, temporally distorting the pulse. These difficulties could potentially be overcome through the use of a noncollinear geometry, whereby the pump and probe are derived from one interferometer and multiple control pulses from another. While the requirement for tight focusing may make noncollinear propagation difficult, the necessity for the detector to image a restricted section of the focal volume implies that even tight (tens of micron) waists could be overlapped at angles of tens of degrees and still form a common interaction region, and could be improved with diffractive shaping of the foci.

The result of applying dual control pulses (C1 and C2) to the vibrational wavepacket in $D_2^+$ is presented in Fig. 8. The delay between the pump ($t = 0$) and probe is scanned for $0 < (C1, C2) < 200$ fs in 1 fs steps. A pump intensity of $10^{14}$ Wcm$^{-2}$ is employed to launch the wavepacket and a pump–control intensity ratio of 3:1 is defined for both C1 and C2. To elucidate the transfer of population, the most populated vibrational state as a function of C1, C2 delay is presented as a colour map in Fig. 8a. A regular modulation of the most populated state is predicted: at small temporal separations from $t = 0$, vibrational states up to $v = 9$ are populated with a periodicity defined by the average period of oscillation of the $D_2^+$ molecular ion. At larger delays at constant C1 or C2 delay, an overlap of the repetitive structure is found, the result of the control pulses acting on a more spatially dispersed wavepacket. This effect is magnified along the C1xC2 diagonal, resulting in a suppressed level of control.

Clearly, the purity of the vibrational population is of interest to coherent control applications and the most populated state is only part of the story, hence the contrast of the population as a function of C1 and C2 is presented in Fig. 8b. The contrast is calculated by taking the difference of the peak population to the average of other populations and calculating the ratio to the total population. Groups of pronounced ridges are found with a similar temporal smearing effect as the C1 and C2 delays are increased from $t = 0$. Interestingly, the maximum contrast only degrades by a small amount as the C1 and C2 delays are increased, rather the ridges blur into each other, which is again a result of the wavepacket dispersing spatially.

Taken in isolation, the maximum populated state or contrast plots are of limited use. By overlaying the two results (Fig. 8c), the relative purity achieved by the dual pulse control scheme is revealed. Around t = 0, a poor control outcome is observed. For C1 or C2 delays up to 100 fs, well isolated islands of optimal control are found, allowing single state access up to $v = 7$ with significant contrast. These regions offer the best chance of experimentally resolving the wavepacket modification. At C1 or C2 delays above 100 fs, the high contrast islands will be difficult to separate, however as the unperturbed wavepacket is known to revive around 580 fs, further investigation is required. Nonequal intensity control pulses may also allow an interesting mix of final states to be populated.

As demonstrated in the single control pulse theoretical and experimental results, variation of the laser intensity has a dramatic influence on the observability of the control operation. In Fig. 9, we demonstrate the same is even more so the case for the dual-control pulse scheme. For a fixed pump intensity of $10^{14}$ $\mathrm{Wcm^{-2}}$, the pump–control intensity is varied from 3:1 (as in Fig. 8) to 2:1 and 1:1 and a subsection of the C1–C2 variation landscape is presented where the clearest manipulation is found in Fig. 8. A pump–control ratio of 2:1 significantly improves the contrast of the final state populations over a ratio of 3:1, particularly for the lowest lying states. This is the result of the more intense control pulse distorting the potential ore severely, thus driving all states including the lowest lying. Despite the increase in control intensity from $3.33 \times 10^{13}$ to $5 \times 10^{13}$ $\mathrm{Wcm^{-2}}$, the high-contrast maximal population ridges do not shift significantly with C1 or C2 delay. As the pump–control ratio is further increased to 1:1 the population and contrast of the $v = 0$ state is observed to be enhanced even further than the 2:1 or 3:1 cases, however the disruption of the ensemble is now so large that all other vibrational states are erratically populated. The loss of the regular structure seen in the 2:1 and 3:1 cases is therefore indicative of the upper useful limit of the dual-control scheme. To further improve the contrast or to populate a pre-defined distribution of states will require additional control pulses.

## 5 Conclusions and outlook

We have demonstrated that controlling the motion of a bound vibrational wavepacket in $D_2^+$ by an ultrashort control pulse can be experimentally implemented and quantified. The redistribution of vibrational population can be recovered using a relatively simple quasi-classical model that incorporates tunnel ionization and dynamic Stark-shift deformation of the potential surface. The simplicity of this model allows it to be integrated over the focal volume, making realistic comparisons with experiment possible. While this model is approximate, it demonstrates the validity of applying a strong-field treatments to a simple system. The sensitivity of the control process to the intensity of all pulses employed is explored, and the requirement for accurate quantification of the focal intensity distribution is established.

A novel application of the quasi-classical model has been presented, allowing a systematic study of the application of two intense few-cycle control pulses. High fidelity population transfer to individual vibrational states is predicted, and following earlier discussions of experimental feasibility, we demonstrate how the range of available states and the transfer contrast depends heavily on the intensity of the control pulses, establishing an upper limit. With access to more computational power, this systematic approach could be improved to search for an optimal outcome to a pre-defined final state using genetic algorithms, with applications in quantum information.[49–51]

The manipulation of a vibrational wavepacket by a strong-field control pulse has very interesting applications to attosecond science. It has recently been demonstrated that the localization of the electron in $D_2^+$ can be externally manipulated by applying a carefully defined light field.[52–54] By varying the relative phase of the carrier and envelope of a few-cycle pulse (referred to as the carrier-envelope phase, CEP), the electron is observed to be driven from one nuclei to the other. This manipulation of the electron wavepacket is evidenced from the asymmetry of the photodissociation or Coulomb explosion process.[55] Such experimental demonstrations naturally point to controlling the vibrational wavepacket while simultaneously driving the electron motion: Calvert *et al.*[56] theoretically showed that significant asymmetry should be observable while modifying the vibrational population. Such methods allowing the nuclear and electronic motions to be selectively directed, allowing additional coherent control routes for strong-field science.

Rather than using the controlled light field of a few-cycle pulse, it has also been shown recently that attosecond XUV pulses can be employed to launch vibrational wavepackets in $D_2^+$ , improving the temporal resolution by around an order of magnitude.[57] Launching the wavepacket by single-photon photoionization rather than tunnel ionization allows the populated rovibrational and electronic states to be selected to some degree. Attosecond pump–probe methods will make a significant contribution to coherent control, as coupling the electronic and nuclear motions allow dynamic wavepacket manipulation beyond the Born–Oppenheimer approximation.

## Acknowledgements

This work was supported by the Engineering and Physical Sciences Research Council (EPSRC) and the Science and Technology Facilities Council (STFC), UK. CRC and RBK acknowledge financial support from the Department of Education and Learning (NI). We wish to thank Chris Froud, Edmond Turcu, Cephise Cacho and Emma Springate of the Artemis Laser Facility, STFC Rutherford Appleton Laboratory, UK for operating the laser system employed to generate the experimental measurements in this work.

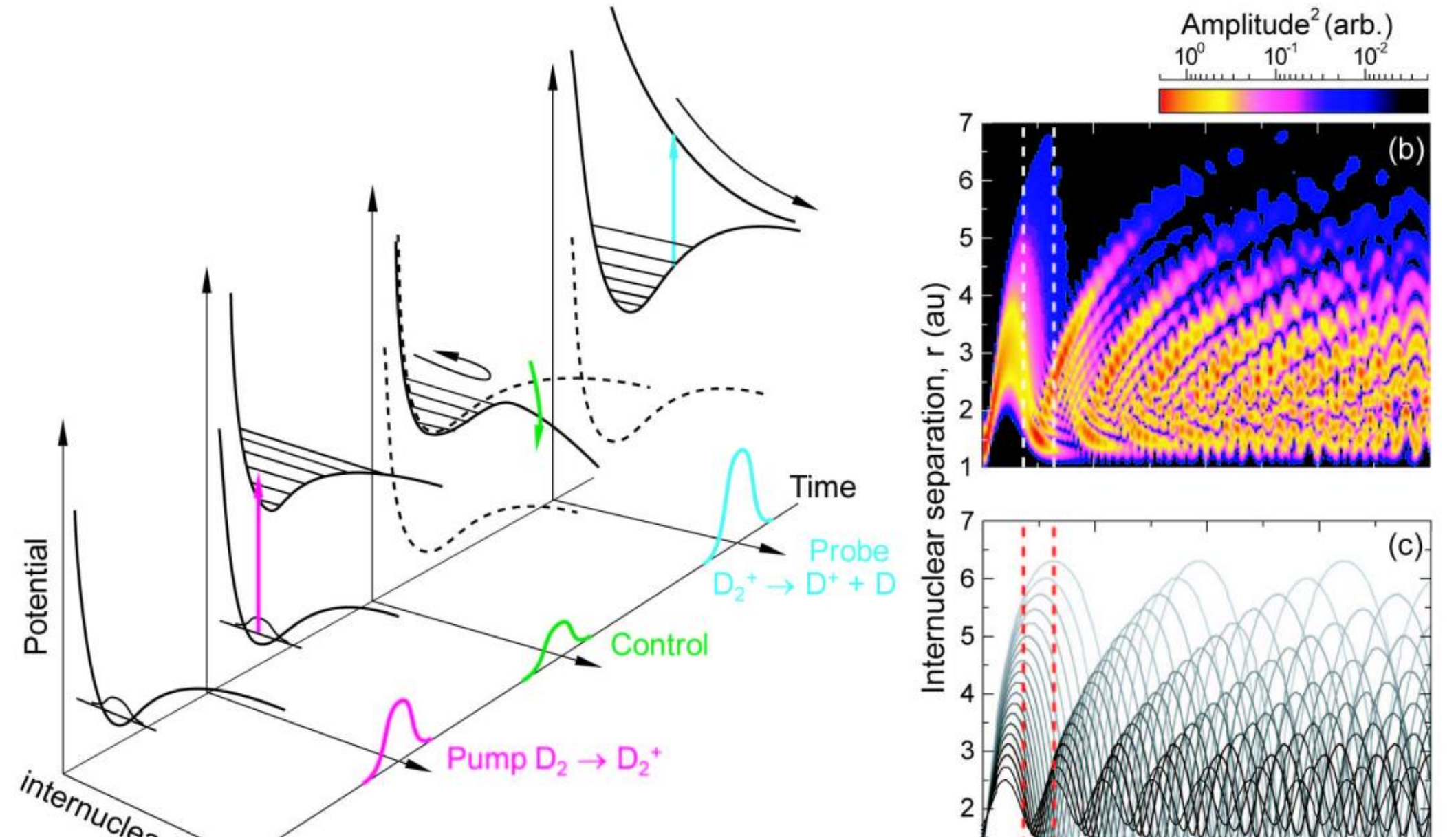

**Fig. 1** Schematic of the pump–control–probe scheme and unperturbed vibrational wavepacket motion. (a) The pump pulse (intensity $4 \times 10^{13}$ Wcm$^{-2}$ to $1.1 \times 10^{14}$ Wcm$^{-2}$) ionizes at room temperature $D_2 \rightarrow D_2^+$ creating a coherent superposition of states. Some time later the control pulse (intensity $1.3 \times 10^{13}$ Wcm$^{-2}$ to $3.7 \times 10^{13}$ Wcm$^{-2}$) distorts the molecular potential energy surface via the polarization of the electronic orbital by the laser field, causing the wavepacket to rapidly adjust. The population redistribution and phase shift caused by the control pulse is imaged by photodissociating the $D_2^+ \rightarrow D^+ + D$ in the probe pulse (intensity $6 \times 10^{13}$ Wcm$^{-2}$ to $2 \times 10^{14}$ Wcm$^{-2}$). (Right) The unperturbed vibrational wavepacket created in the $D_2^+$ molecule: the solution to the time-dependent Schrodinger equation (top, b) and current quasi-classical model[41] (bottom, c) are compared. Both exhibit the same return to the inner turning point, the region over which the control pulse is applied is indicated by vertical dashed lines.

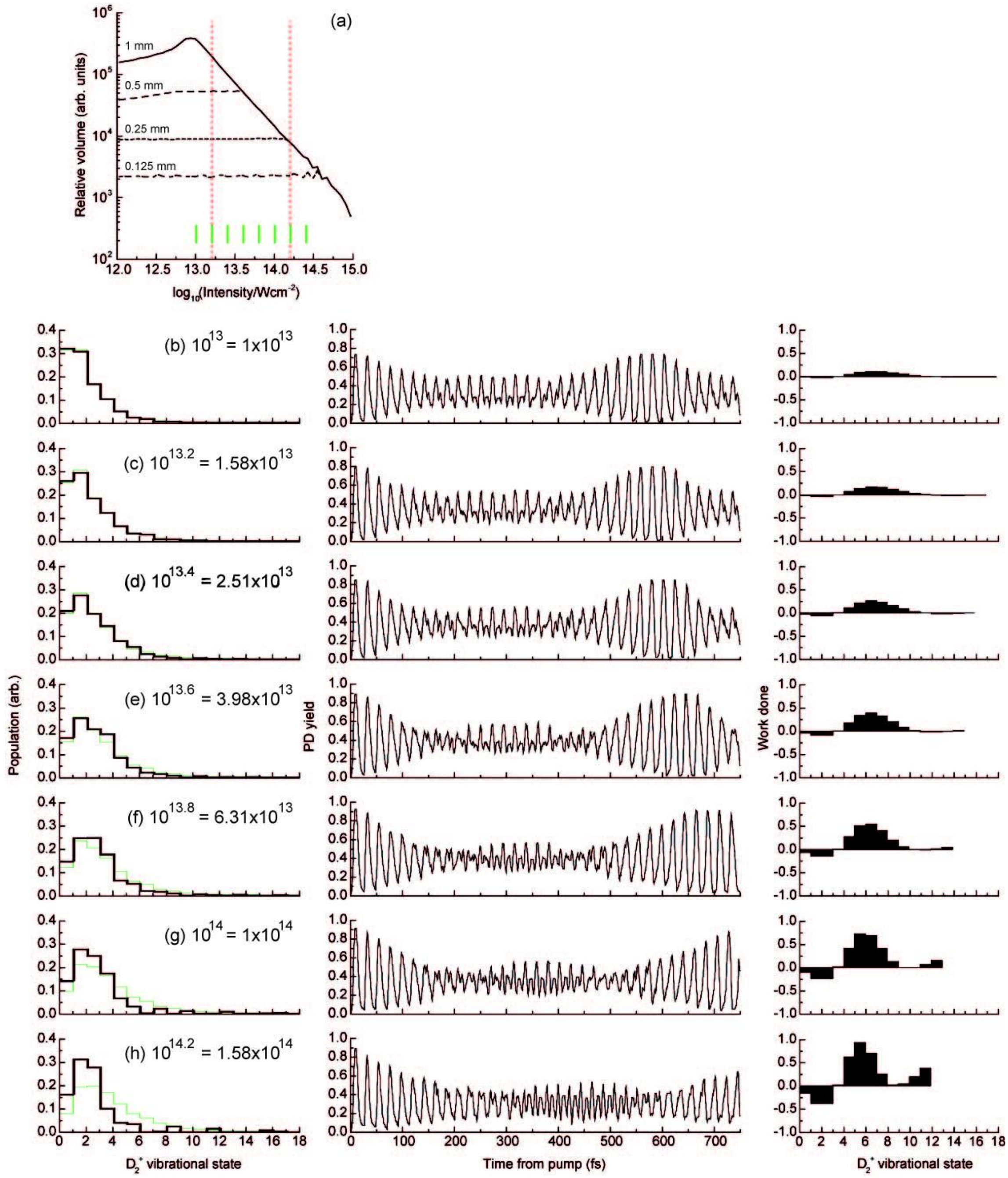

**Fig. 2** (a) The focal volume of a typical ultrafast laser focus ($\lambda$ = 790 nm, focal length $f$ = 50 mm, 1/e$^2$ beam radius D = 1.5 mm, total propagation distance from source to focal volume L = 5 m) characterized with a relative-volume histogram. The distribution of intensity is found by numerically solving the Huygens–Fresnel diffraction integral. By reducing the radius of the circular aperture through which the PD fragments are detected, a subsection of the focal volume is imaged. The red dashed lines indicate the intensity range over which stable $D_2^+$ ions are produced, and the green ticks correspond intensities for which the calculations are provided below. (b) To (h) demonstration of the QCM for a pump–control intensity ratio of 3:1, where the pump, control and probe have a duration of 6 fs. (Left) Initial (green line) and final (black line) vibrational population distributions for a varying pump pulse intensity. (Centre) Photodissociation yield as a function of pump–probe delay. (Right) Work done on or by the vibrational wavepacket as a function of vibrational state. The modification of the vibrational population can be seen to be the result of significant work done.

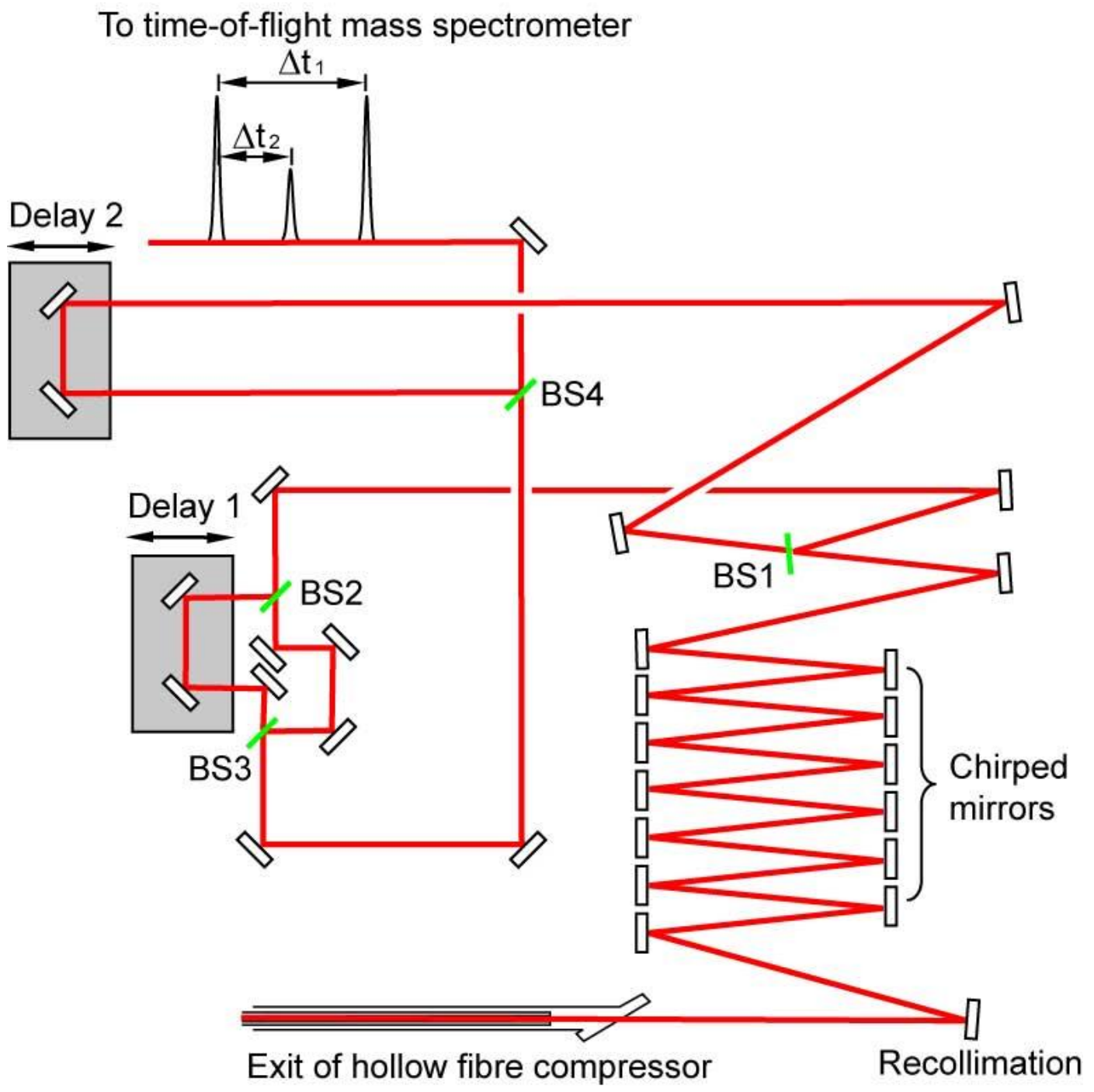

**Fig. 3** Schematic of the ultrafast pulse manipulation allowing the observation of strong-field vibrational wavepacket control. Two nested Mach–Zehnder interferometers constructed using low dispersion silver mirrors and thin dielectric beamsplitters produce the pump–control–probe sequence with independently variable delays.

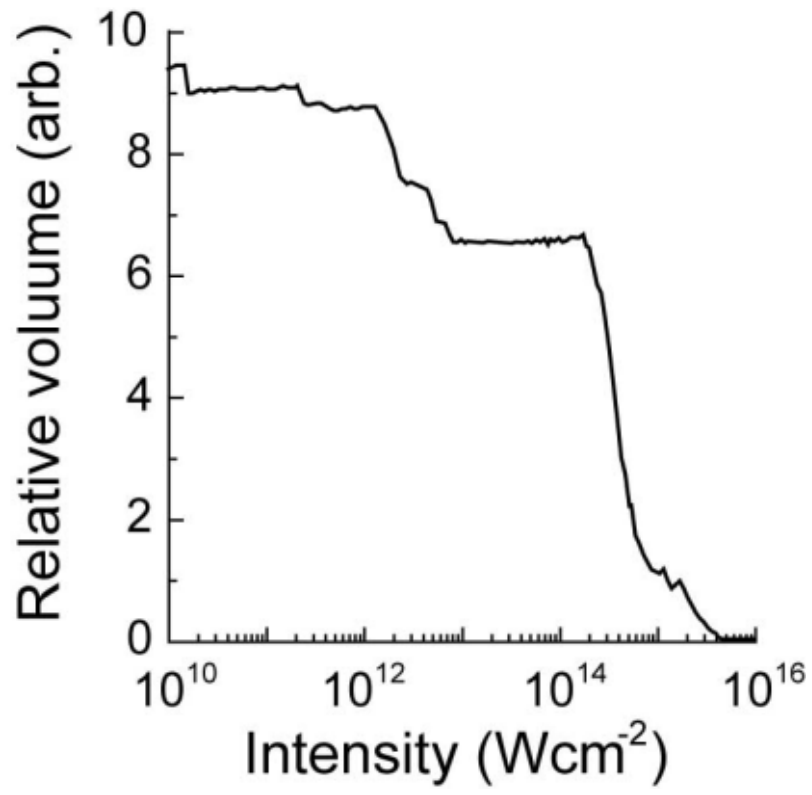

**Fig. 4** Focal histogram for experimental demonstration.

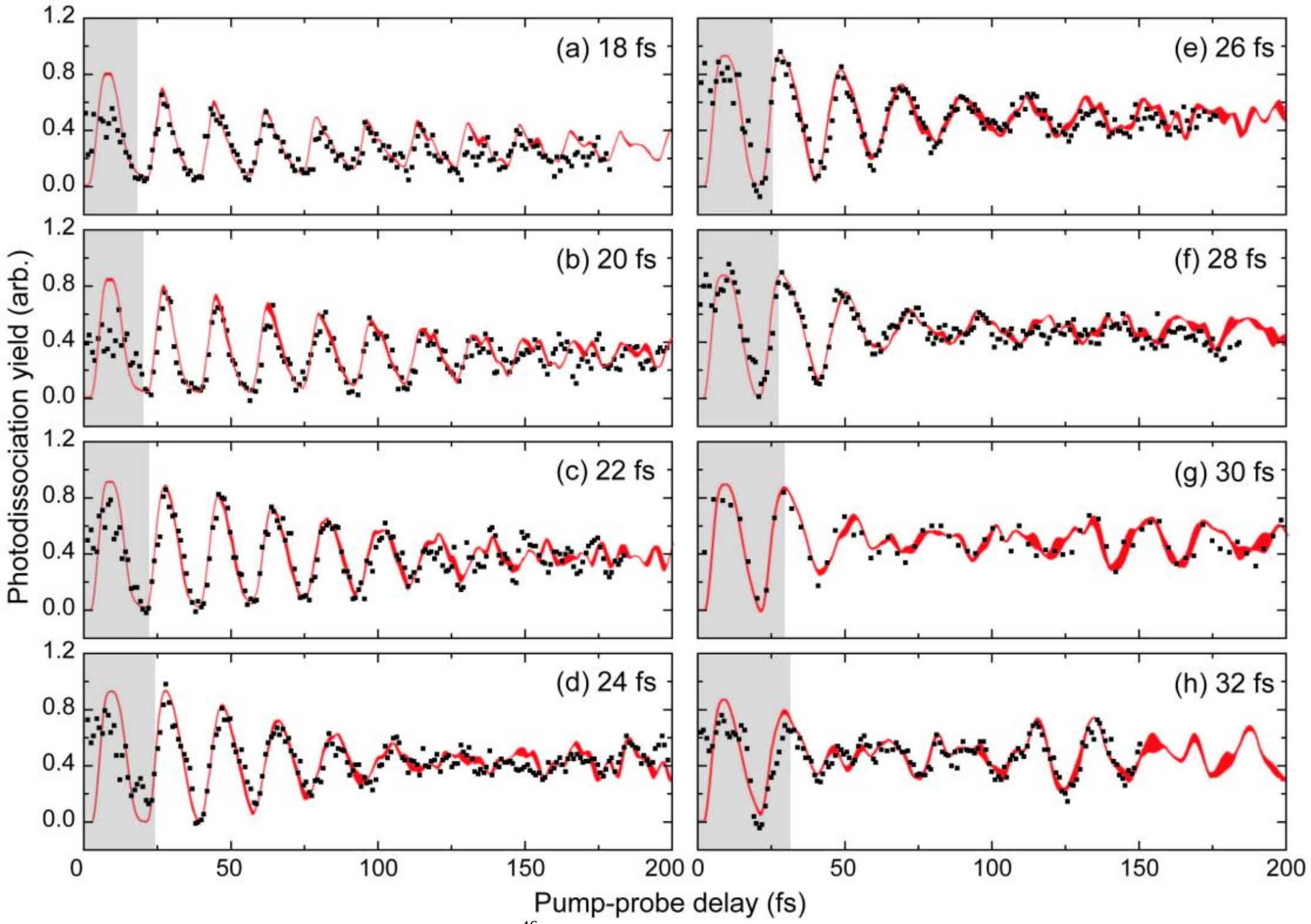

**Fig. 5** Experimentally measured photodissociation yield[46] and predicted focal–volume–integrated yield derived from the quasi-classical model.[41] The agreement between theory (red line) and experiment (black squares) demonstrates that the control pulse is manipulating the wavepacket evolution in a quantifiable manner and the molecular PES returns to the field free state following the control pulse. The varying vertical thickness of the predicted yield indicates the uncertainty in fitting the experimental results. The grey shading indicates the presence of the control pulse, hence the QCM is not expected to describe the experimental data accurately in this region.

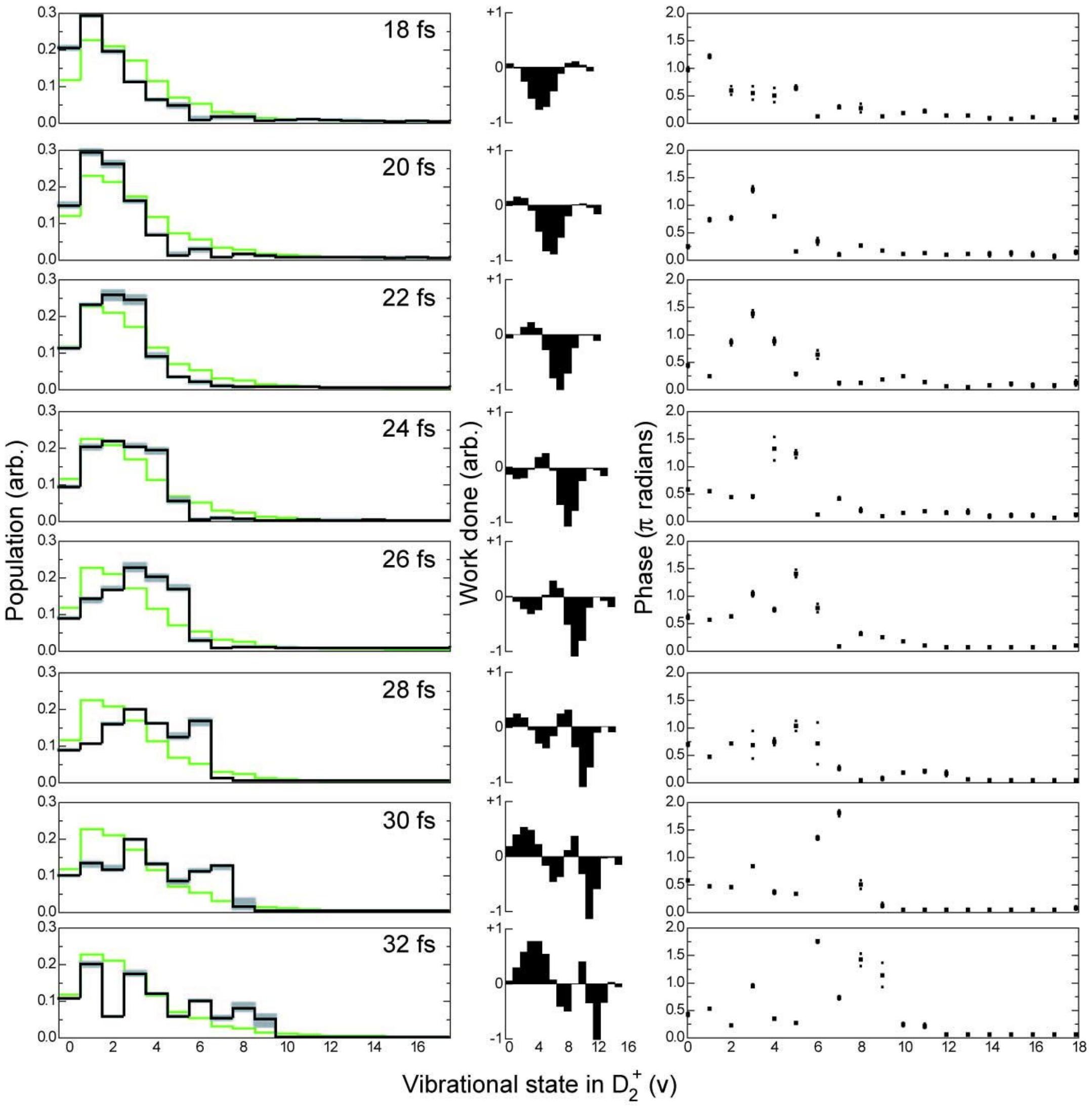

**Fig. 6** The best-fit predictions of vibrational population, work done on or by the control pulse, and classical phase distributions as the pump–control delay is varied. As the control delay is sequentially changed from 18 to 32 fs, a significant shifting in the vibrational population (black) with respect to the initial distribution of states (green) is observed, and is a result of the work done. The largest change in classical phase follows the maximal population shift. The uncertainty in population (grey bars) and classical phase (small markers) is derived from the uncertainty when fitting the experimental PD yield (Fig. 5).

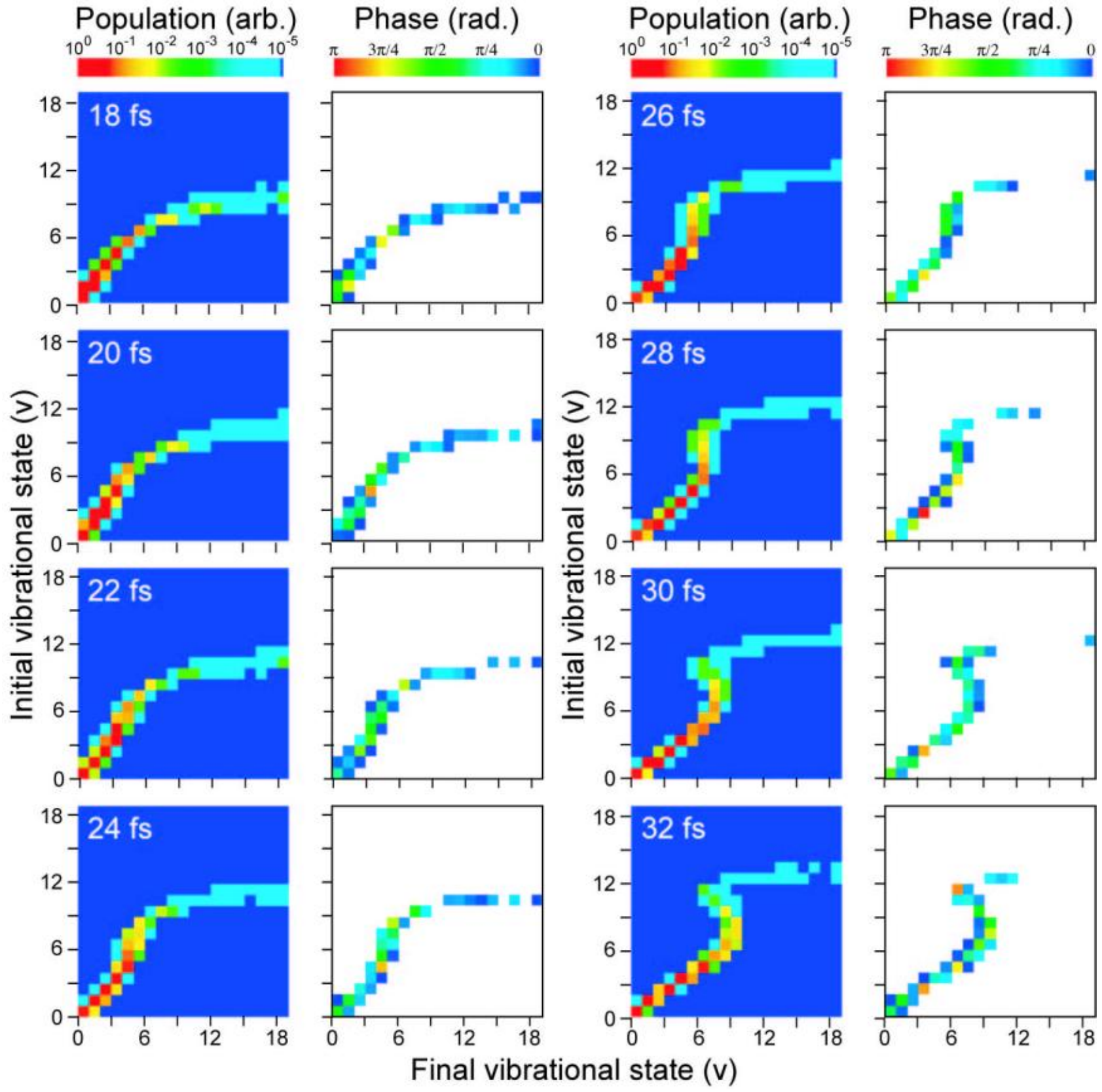

**Fig. 7** Volume-integrated vibrational population and phase matrices as the pump–control delay is scanned. The QCM compares all initial and final trajectories allowing arbitrary transfer of population. The amplitude of the initial state is defined by the tunnel ionization step; when a match to the trajectory of the final state is found, this population is transferred to the final state. To facilitate a best-fit, it is necessary to include a phase offset, summarized in the phase matrix. This process is repeated across the focal histogram then summed vertically to produce Fig. 6.

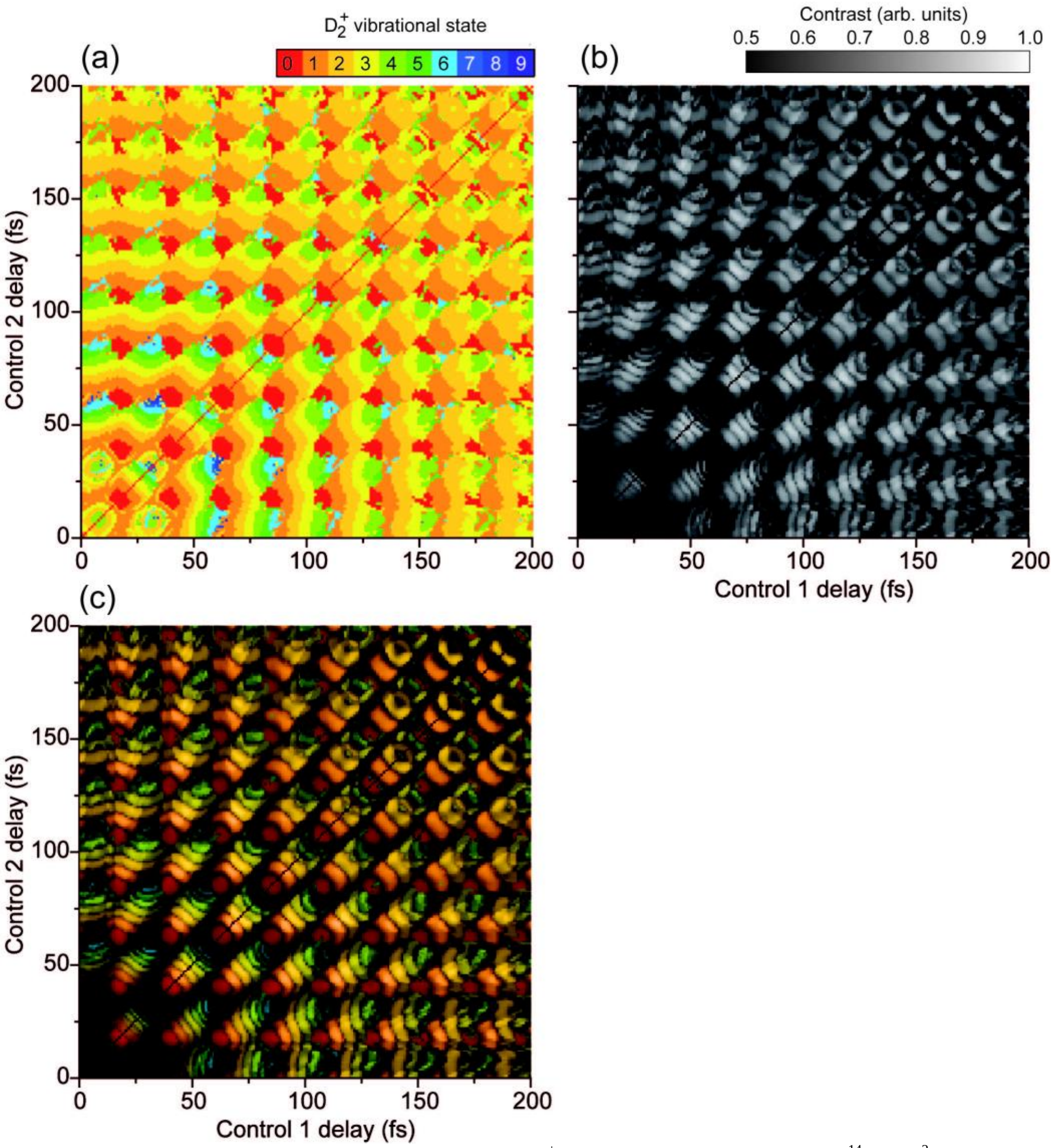

**Fig. 8** Modelling the outcome of applying two control pulses to the $D_2^+$ ensemble. Pump intensity = $10^{14}$ Wcm$^{-2}$, pump–control intensity ratio = 3:1, duration of pump, control and probe = 6 fs. (a) Colour map of the most populated vibrational state as control pulses C1 and C2 are scanned from 0 to 200 fs. As pulses C1 and C2 are identical, the region below the diagonal is reflected. (b) Vibrational state contrast, C as a greyscale map, where C = ($pop_{max}$ _ $pop_{min}$)/ ($pop_{max}$ + $pop_{min}$), and $pop_{min}$ is the mean of the remaining populations not equal to $pop_{max}$. (c) Most populated state and contrast map overlaid to illustrate the final state fidelity that can be achieved, so the ridged colour indicates vibrational state and the luminosity indicates the purity of state.

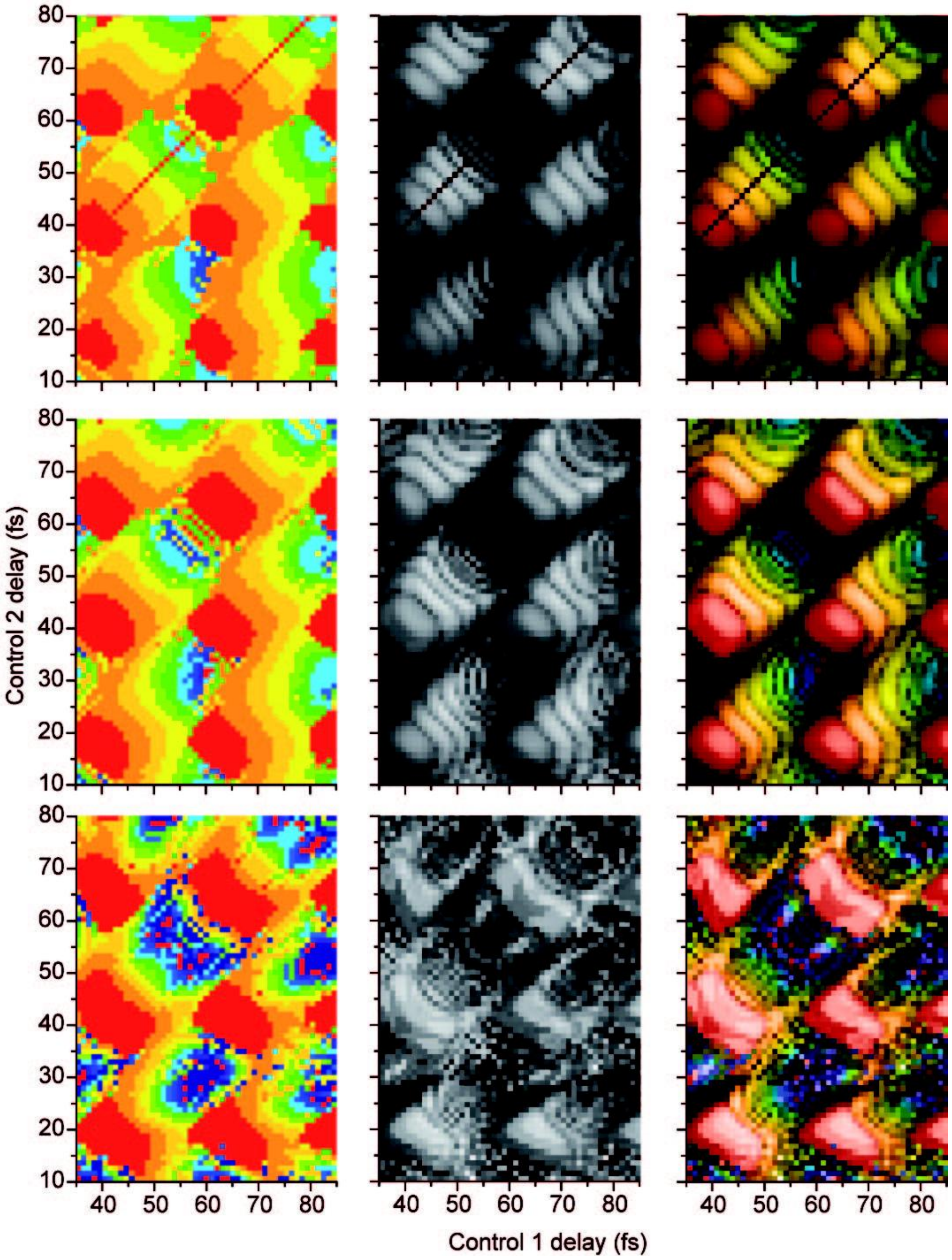

**Fig. 9** Section of the two control pulse landscape as the pump–control ratio is varied from 3:1 (top), 2:1 (middle) and 1:1 (bottom). The pump intensity = $10^{14}$ Wcm$^{-2}$ and the duration of pump, control and probe = 6 fs. The most populated vibrational state, contrast and final state fidelity as in Fig. 8.